\documentclass{PoS}
\usepackage{amsmath}

\title{
\vspace*{-1.5cm}
{\normalsize {\rm \hfill{ LLNL-PROC-68031} }} \\
\vspace*{1.5cm}
Walking and conformal dynamics in many-flavor QCD}

\ShortTitle{Walking and conformal dynamics in many-flavor QCD}

\author{
Yasumichi Aoki$^a$, 
Tatsumi Aoyama$^a$, 
Ed Bennett$^b$, 
Masafumi Kurachi$^c$, \quad \quad
Toshihide Maskawa$^a$,  
Kohtaroh Miura$^d$, 
Kei-ichi Nagai$^a$, 
\speaker{Hiroshi Ohki}$^e$, \quad \quad \quad
Enrico Rinaldi$^f$, 
Akihiro Shibata$^g$, 
Koichi Yamawaki$^a$, 
and Takeshi Yamazaki$^h$ \\
\centering{(LatKMI Collaboration)} \\

\begin{flushleft}
$^a$ 
Kobayashi-Maskawa Institute for the Origin of Particles and the Universe (KMI), 
Nagoya University, Nagoya, 464-8602, Japan \\
$^b$ 
Department of Physics, Swansea University, Singleton Park, 
Swansea SA2 8PP, UK \\
$^c$ 
Institute of Particle and Nuclear studies, High Energy Accelerator Research
Organization (KEK), 
Tsukuba 305-0801, Japan \\
$^d$ 
Centre de Physique Theorique(CPT),
Aix-Marseille Univerisity, Campus de Luminy, Case 907,
163 Avenue de Luminy, 13288 Marseille cedex 9, France \\
$^e$ 
RIKEN BNL Research center, Brookhaven National Laboratory, 
Upton, NY, 11973, USA \\ 
$^f$ 
Lawrence Livermore National Laboratory, Livermore, 
California, 94550, USA \\
$^g$ 
Computing Research Center, High Energy Accelerator Research Organization (KEK),
Tsukuba 305-0801, Japan \\
$^h$ 
Graduate School of Pure and Applied Sciences,
University of Tsukuba, Tsukuba, Ibaraki 305-8571, Japan \\
\end{flushleft}

}

\abstract{
In the search for a realistic walking technicolor model,
QCD with many flavors is an attractive candidate.
From the series of studies by the LatKMI collaboration,
we present updated results of the scaling properties
of various hadron spectra, including the (pseudo)scalar, vector,
and baryon channels, for $N_f=8$ QCD analyzed with the HISQ action.
By comparing these with $N_f=12$ QCD, which has properties consistent
with conformality, possible signals of walking dynamics are discussed.
We also present a preliminary result of the flavor-singlet pseudoscalar 
mass in many-flavor QCD.
}

\FullConference{The 33rd International Symposium on Lattice Field Theory\\
		14 -18 July 2015\\
		Kobe International Conference Center, Kobe, Japan*}

\begin{document}

\section{Introduction}
\vspace{-2mm}
SU(3) gauge theory with many flavors  is 
a very good candidate for a walking technicolor model.
The LatKMI collaboration has been systematically investigating  
the SU(3) gauge theory with $N_f$ fundamental 
fermions with $N_f$= (0), 4, 8, 12, and 16 
using a common setup of the lattice action.
We utilize the Highly improved staggered quark (HISQ) 
action with tree-level Symanzik gauge action (HISQ/tree).
Our previous results suggested that $N_f=8$ QCD could 
have a walking behavior~\cite{Aoki:2013xza}. 
Similar results were also given in \cite{Appelquist:2014zsa}.
More interestingly,  the flavor-singlet scalar mass 
is found to be as light as the Nambu-Goldstone (NG) pion ($\pi$)
in $N_f=8$ QCD~\cite{Aoki:2014oha}.
In this proceeding, 
we present our updated results of the scaling 
properties of various hadron spectra, including 
the pseudoscalar mass ($M_\pi$), 
decay constant ($F_\pi$), vector mass ($M_\rho$), 
and nucleon mass ($M_N$) in comparison with $N_f=12$ QCD.
We also present a new result of the measurement of the flavor-singlet 
pseudoscalar ($\eta'$) mass for the first time in the many flavor QCD.
The mass of $\eta'$ meson is interesting, 
since the fermion loop contribution, 
which would naturally  enhance as $N_f$ increases, 
plays an essential role. 
Using a topological charge density operator and the gradient flow 
we can obtain a good signal for the $\eta'$ meson two-point function. 
All the updated results shown here are preliminary.

\vspace{-3mm}
\section{Simulation setup and simple analysis}
\vspace{-2mm}
We have been generating configurations at $\beta=3.8$ 
with lattice volumes $(L, T)=(18, 24)$, $(24, 32)$, $(30, 40)$, $(36, 48)$
and $(42, 56)$,
for various fermion masses.
Compared to our previous results in Ref.~\cite{Aoki:2013xza}, 
we have added new simulation points in the smaller mass region  
of $m_f= 0.012$ and $0.015$ on $L=42$  
with $2,200$ and $4,760$ HMC trajectories,
and accumulated more configurations at smaller masses on larger volumes. 
We have now typically ten times many trajectories than the previous data
for smaller mass region.
The details of simulation parameters and 
updated results can be found in Ref.~\cite{nf8}, 
where we should mention that some spectrum data have been changed
at the $1 \sigma$ level from the previous results.
This is due to the fact that there are unexpectedly long auto correlation lengths 
for HMC history, 
which appeared also in the topological history 
presented in Ref.~\cite{topology}. 
In the present analysis,
taking a longer HMC trajectory with smaller fermion masses, 
we obtain more reliable results, 
which enable us to deeply investigate the scaling behavior of the 
various hadron spectra. 
As shown later, this improvement has affected the numerical result for 
the finite-size hyperscaling analysis, 
while the statement that there exists an (approximate) conformal behavior is unchanged.

\begin{figure}[h]
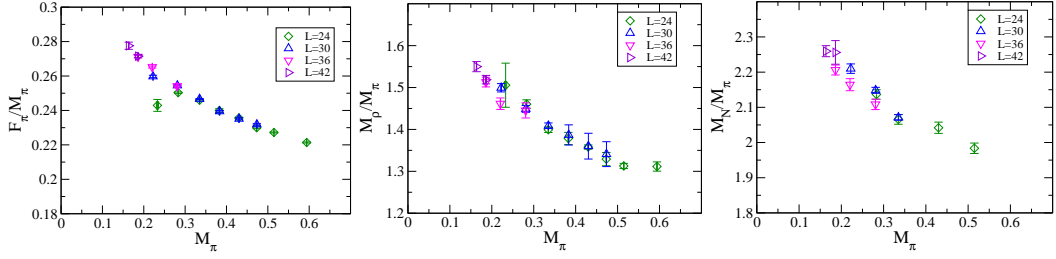

\begin{center}
\includegraphics[width=1.8in,clip]{fpi-mpi.eps}
\includegraphics[width=1.8in,clip]{mpv-mpi.eps}
\includegraphics[width=1.8in,clip]{mN-mpi.eps}
\end{center}
\vspace{-8mm}
\caption{
$F_\pi/M_\pi$ v.s. $M_\pi$ (Left), 
$M_\rho/M_\pi$ v.s. $M_\pi$ (Center),
and $M_N/M_\pi$ v.s. $M_\pi$ (Right) 
for $N_f=8$.
}
\label{fig:ratio}
\end{figure}

As a simple analysis,  we study dimension-less ratios of the 
physical quantities as $M_\rho/M_\pi$, $F_\pi/M_\pi$,
and  $M_N/M_\pi$ as a function of $M_\pi$ shown in Fig.~\ref{fig:ratio}.
Those ratios are increasing towards the chiral limit. 
A similar tendency can be seen in our $N_f=4$ data, 
and is clearly different from the one in $N_f=12$, 
where we find those ratios have mild $M_\pi$ dependence,
and become a constant in the small-$M_\pi$ region.
As for the ratio analysis, 
our updated result is consistent with the previous result.

\vspace{-3mm}
\section{Hyperscaling analysis}
\vspace{-2mm}

Although the ratio analysis shows a consistent behavior with the 
theory having broken chiral symmetry, 
we can still expect some remnant of the conformal symmetry,  
if the theory is near the edge of the conformal window.
In fact, in our previous data we have found that 
each quantity has an approximate hyperscaling in the intermediate mass region.
As shown below, using the updated result with smaller fermion masses, 
it turns out that this scaling can be seen even at smaller fermion masses.
This is one major change in our updated result.

\begin{figure}[h]
\begin{minipage}[b]{.5\textwidth}
\begin{center}
\includegraphics[width=1.8in,clip]{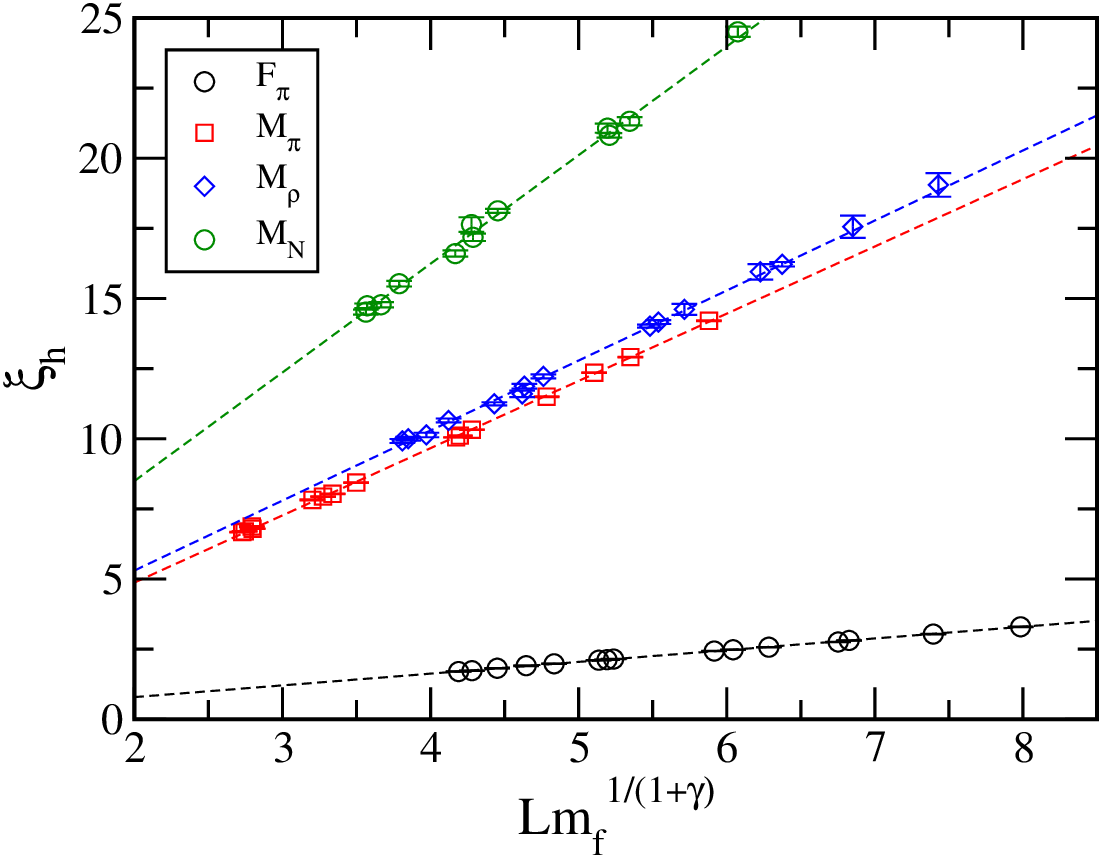}
\end{center}
\vspace{-5mm}
\caption{
Individual finite-size hyperscaling fit for each $\xi_h$.
}
\vspace{-2mm}
\label{fig:naive}
\end{minipage}
\quad 
\begin{minipage}[b]{.5\textwidth}
\makeatletter
\def\@captype{table}
\makeatother
\small
\begin{center}
\begin{tabular}{c| c | c} 
& $\gamma$ & $\chi^2/\mathrm{dof}$  \\ \hline
$F_\pi$ & 1.010(6) & 1.7 \\
$M_\pi$ & 0.631(3) & 19.2 \\
$M_\rho$ & 0.904(18)  & 1.7 \\
$M_N$  &  0.838(20) & 3.0  \\  
\end{tabular}
\end{center}
\caption{
Result of the naive finite-size hyperscaling fits. 
}\label{tab:naive}
\end{minipage}
\end{figure}

We carry out an individual finite-size hyperscaling fit 
using a naive function,  
\vspace{-2mm}
\begin{eqnarray}
\xi_h = c_0 + c_1 x, 
\label{eq:naive}
\vspace{-2mm}
\end{eqnarray}
where 
$\xi_h = L M_h$, $h=\pi, \rho,$ and $N$, 
or $ \xi_F = L F_\pi$,
and  $x=L m_f^{1/(1+\gamma)}$.
In the finite-size hyperscaling analysis 
we only use the updated data
that covers a simulation parameter region 
with $0.012 \le m_f \le 0.08$ and 
$24 \le L \le 42$~\footnote{A full analysis including the previous data with heavier mass region will be
presented in Ref.~\cite{nf8}.}.
The fit results are shown in Fig.~\ref{fig:naive}
and Table~\ref{tab:naive}.
Unlike the previous result  (c.f. Table V in Ref.~\cite{Aoki:2013xza}), 
the naive fit works even in a region of smaller masses 
except for $M_\pi$.
The resulting value of $\gamma$ is $\mathcal{O}(1)$, 
but it is not universal. 

\begin{figure}[h]
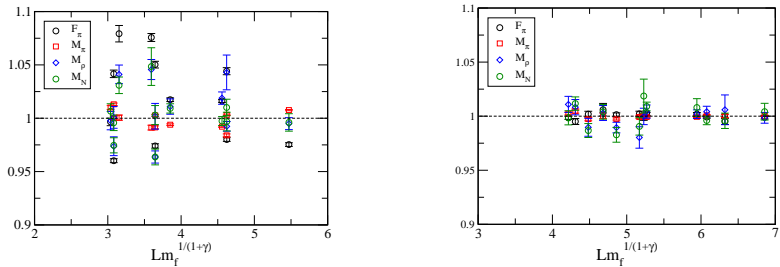

\begin{center}
\includegraphics[width=1.7in,clip]{global.eps}
\quad \quad \quad \quad
\includegraphics[width=1.7in,clip]{omega.eps}
\vspace{-5mm}
\end{center}
\caption{
Simultaneous finite-size hyperscaling fit
for the naive function (Left), and with correction (Right).
}
\label{fig:global}
\end{figure}

From the mass-deformed conformal theory point of view, 
non-universality and a large $\chi^2/\rm{dof}$
might be caused by corrections to hyperscaling.
To test such a possibility, we carry out a finite-size hyperscaling with mass corrections.
Among various types of the mass corrections, 
we adopt a renormalization group inspired correction term~\cite{Cheng:2013xha}
as a benchmark test, which is
\begin{eqnarray}
\frac{\xi_h}{1+c_2 m^\omega} = c_0 + c_1 x.
\label{eq:correction}
\end{eqnarray}
In this formula, there is another exponent $\omega$ in the correction term, 
whose theoretical origin comes down to the critical exponent of the irrelevant operator 
$g$ (gauge coupling) in the vicinity of the infrared fixed point.
As $\omega$ can not be analytically determined in the present analysis, 
we treat it as a fit parameter. 
Then we carry out the simultaneous finite-size hyperscaling fit for 
the quantities of $F_\pi$, $M_\pi$, $M_\rho$, and $M_N$ with 
common values of $\gamma$ and $\omega$.
As a comparison we also perform a simultaneous fit 
without a correction (Eq.~\ref{eq:naive}).
Both results 
are shown in Fig.~\ref{fig:global},
where the vertical axis $y$ in the $x$-$y$ plane means,
\vspace{-2mm}
\begin{eqnarray}
y = 
\begin{cases}
\frac{\left(\xi_h-c_0\right)/c_1}{x} & \text{(naive fit)},
\\ 
\frac{\left(\xi_h/(1+c_2 m^\omega) -c_0\right)/c_1}{x} & \text{(with correction)}.
\end{cases}
\end{eqnarray}
The data are distributed
around the fit line, {\it i.e.} $y=1$, 
where the fitted data with mass correction 
are closer to the fit line than the naive one.
The fit results are tabulated in Table~\ref{tab:correction}.
As a result, 
the mass correction term improves the fit accuracy and 
we obtain a reasonable $\chi^2/\rm{dof}$ 
and $\gamma \sim 1$. 

However, we should note that
in this approach 
there is no systematic way 
to incorporate finite mass and volume corrections to the universal hyperscaling relation, 
and the value of $\gamma$ 
depends on the model in general.
In addition, the contribution of the correction term for $M_\pi$ 
is found to be comparable to that of the naive hyperscaling term in the simulation mass region.
Note that $c_2$ differs depending on the quantity.
In this situation, 
it is not immediately obvious whether each data actually shows a universal scaling 
towards the chiral limit.
(As shown later, there is a clear difference in the correction
between $N_f=8$ and $12$.)
Even if this is the case, 
$\gamma(M_\pi) \sim 1$ is also consistent with the chiral broken phase, 
since it coincides with the leading order of $m_f$ dependence of $M_\pi$ 
in the chiral perturbation theory (ChPT) 
formula~\footnote{
In fact, a ChPT-like fit also works in our data, 
where we can obtain a tiny non-zero $F_\pi$ in the chiral limit, 
and the higher order term for $M_\pi^2$ is 
required to fit the data~\cite{nf8}. }.
In either case whether $N_f=8$ QCD 
is in the conformal phase with $\gamma \sim 1$
or in the chirally broken phase, 
we can expect that $M_\pi$ behaves like $M_\pi \sim m_f^{1/2}$ 
as approaching the chiral limit.

In the next section, 
in order to see such a scaling behavior, 
we study mass (scale) dependence of $\gamma$ for each quantity in detail.

\begin{table}[!h]
\begin{center}
\small
\begin{tabular}{c| c | c | c } 
& $\gamma$ & $\omega$ & $\chi^2/\rm{dof}$  \\ \hline
naive fit & 0.708(3) & $-$ & 84.2 \\ 
with correction 
& 1.02(4) & 0.35(2) & 2.3 
\end{tabular}
\vspace{-5mm}
\end{center}
\caption{
Fit result for simultaneous finite-size hyperscaling 
with and without a mass correction term.
}\label{tab:correction}
\end{table}
\vspace{-4mm}

\vspace{-4mm}

\section{Prospect towards the chiral limit}

To see a scaling property in detail, 
we need to systematically study the fermion mass dependence of 
$\gamma$. 
We consider a window for the fermion mass parameter
which corresponds to a range for three sequential fermion masses, 
and slide it from $[m_f^{\rm min}, m_f^{\rm max}]=
[0.012, 0.02]$ to $[0.07, 0.1]$.
We then perform a fit for each window. 
The data on the largest volume are used for each mass, 
so that we use the naive hyperscaling function $M_h \propto m_f^{1/(1+\gamma)}$. 
An effective mass anomalous dimension (denoted $\gamma_{\rm eff}(m_f))$ 
is calculated as a fit result for each window. 

\begin{figure}[h]
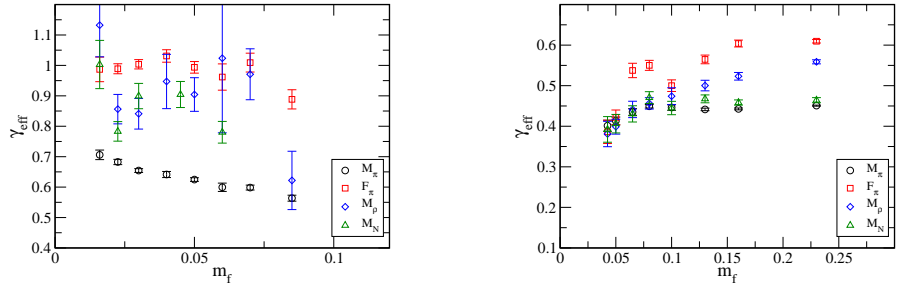

\begin{center}
\includegraphics[width=2in,clip]{geff_nf8.eps} 
\quad \quad \quad \quad
\includegraphics[width=2in,clip]{geff_nf12.eps}
\end{center}
\vspace{-7mm}
\caption{
$\gamma_{\rm eff}(m_f)$  for $N_f=8$ (Left), 
and for $N_f=12$ (Right). 
}
\label{fig:geff}
\end{figure}

The result for $\gamma_{\rm eff}(m_f)$ is plotted 
in the left panel of Fig.~\ref{fig:geff}, 
where the $x$-axis means the central value of the fit range.
We find that the value of $\gamma_{\rm eff}$ for $M_\pi$
increases and it looks like approaching $\sim 1$.
While 
our data are far away from the chiral limit, 
this tendency could be a promising signal for the chirally broken phase. 
In fact, this result is in sharp contrast to the $N_f=12$ result, 
which is shown in the right panel of Fig.~\ref{fig:geff}.
In $N_f=12$, $\gamma_{\rm eff}$ for $M_\pi$ 
never increases towards the chiral limit. 
Furthermore, $\gamma_{\rm eff}$ from various quantities tend to become universal 
in a range of small mass within our statistical accuracy, 
indicating that the system is in the scaling region.
This is consistent with the conformal nature for $N_f=12$ QCD.
A large correction to $M_\pi$ for $N_f=8$ found in the previous section could be
understood 
since it is not in the scaling region, 
while the correction for $N_f=12$ would become important only 
when the data outside of the scaling region were included. 
Although it is obvious 
within the current data 
we can not discriminate both possibilities 
between the strongly coupled conformal theory ($\gamma \sim 1$) 
and the chiral broken theory with approximate hyperscaling,
the tendency of $\gamma_{\rm eff}$ 
found in $N_f=8$ 
is indicative of a walking gauge theory.

\vspace{-3mm}

\subsection{Effective $\gamma$ from Dirac eigenvalues}
\begin{figure}[h]
\begin{center}
\includegraphics[width=1.8in,clip]{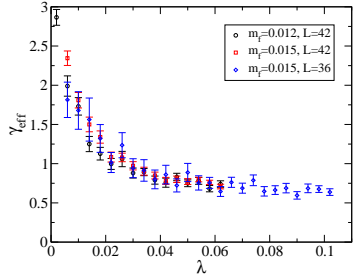}
\end{center}
\vspace{-7mm}
\caption{
$\gamma_{\rm eff}(\lambda)$ from Dirac eigenvalue spectrum
for $N_f=8$.}
\label{fig:evs}
\end{figure}

Another way to calculate an effective $\gamma$ is to 
use the spectrum of the Dirac eigenvalues. 
From the density of eigenvalues, $\rho(\lambda)$, 
with Dirac eigenvalue $\lambda$, 
the scaling law of $\rho(\lambda)$ is given 
as $\rho(\lambda)  \propto \lambda^{\frac{3-\gamma}{1+\gamma}}$.
Thus we can obtain a scale-dependent 
mass anomalous dimension~\cite{Cheng:2013eu}. 
We define an effective anomalous dimension from Dirac eigenvalues 
as 
\begin{eqnarray}
\frac{3-\gamma_{\rm eff}(\lambda)}{1+\gamma_{\rm eff}(\lambda)}
= \frac{\ln{\rho(\lambda+\Delta) - \ln{\rho(\lambda)}}}{\ln{(\lambda+\Delta)-\ln{\lambda}}}.
\label{eq:rho}
\end{eqnarray}
We show the result for the smallest two fermion masses 
with $\Delta = 0.004$
in Fig.~\ref{fig:evs}.
We find $\gamma_{\rm eff} (\lambda \sim 0) \sim 3$, 
which is consistent with a non-zero chiral condensate 
suffering from 
a non-zero fermion mass effect.
Looking at small $\lambda$ ($> m_f$), 
we estimate $0.5 < \gamma_{\rm eff}(\lambda) \le 1$ for $0.03 \le \lambda \le 0.1$.
This is roughly consistent with the one obtained from hadron spectra,
but it requires a more careful study. 

\section{Flavor-singlet pseudoscalar mass}
\begin{figure}[h]
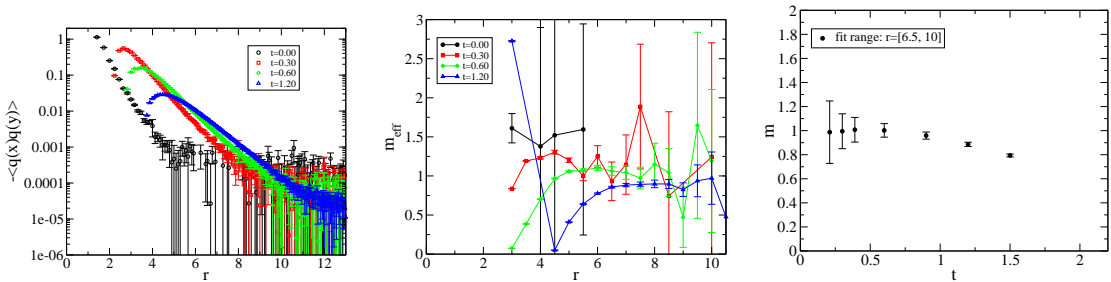

\begin{center}
\includegraphics[width=1.8in,clip]{corr.eps} \quad  
\includegraphics[width=1.8in,clip]{meff_all.eps} \quad  
\includegraphics[width=1.8in,clip]{mfit_dep.eps}
\end{center}
\vspace{-5mm}
\caption{
(Left) Two point correlator for $\eta'$ meson obtained from 
a topological charge density operator for $N_f=8$.
(Center) The effective mass for the $\eta'$ meson. 
(Right) Flow time dependence of the $\eta'$ meson mass.
} 
\label{fig:qq}
\end{figure}

Here we would like to investigate the flavor-singlet pseudoscalar ($\eta'$).
The $\eta'$ meson would be a NG-boson of the axial $U(1)$ symmetry 
of QCD, while its mass is larger compared to the flavor non-singlet pseudoscalar 
($\pi$), which can be attributed to the axial $U(1)$ anomaly,
where the $N_f$ factor manifests in the anomaly contribution. 
The axial anomaly relation tells that 
the topology of QCD can also play an important role in the $\eta'$ meson mass,
so that an investigation of the $N_f$ dependence of the mass
is important to understand QCD. 
We use a topological charge density operator $q(x)$
to calculate the two-point correlation function of the $\eta'$ meson.
We use the gradient flow method~\cite{Luscher:2010iy} 
to improve the statistical accuracy, 
which was already adopted in the measurement of the topological charge and 
susceptibility~\cite{topology}. 
We measure the correlation function $\langle q(x) q(y)\rangle$
for various flow time $t$.
As a preliminary study we calculate the correlation function 
at $m_f=0.02$ on $L=36$ in $N_f=8$~\footnote{
Similar analyses have been done in quenched QCD~\cite{Chowdhury:2014mra} 
and $2+1$ flavor QCD~\cite{Fukaya:2015ara}. }. 
The result for the correlation function 
is shown in the left panel of Fig.~\ref{fig:qq}, 
where $r=|x-y|$.  The mass of the $\eta'$ meson ($M_{\eta'}$) is obtained by 
a fit with $\langle q(x) q(y)\rangle = c K_1(M_{\eta'}r)/r$, where $K_1(x)$ is 
a modified Bessel function and $c$ is a constant.
We estimate an effective mass 
from a fit with range $[r, r+0.5]$ using an asymptotic form
of the above function. 
The result of the effective mass is shown in the middle panel of Fig.~\ref{fig:qq}.
We find that a better plateau is obtained for larger flow time.
To see flow time dependence of the mass, 
we carry out the fit with fixed fit range of $r=6.5-10$.
The result is shown in the right panel of Fig.~\ref{fig:qq}, 
where we find a stable region for $t \leq 1$, 
and in this region a signal becomes better as $t$ increases. 
We quote a mass as $M_{\eta'}=1.00(6)$ at $t=0.6$ for $m_f=0.02$.
We obtain a ratio $M_{\eta'}/M_\rho=3.1(2)$.
This result is much larger than the real-life QCD.
Our result suggests a heavy $\eta'$ meson, 
which might be due to a large fermion loop effect in many-flavor QCD.

\section{Summary}
We have studied the scaling properties of various hadron spectra
in $N_f=8$ QCD.
We found that the ratios of the hadron spectra 
show a similar behavior to the $N_f=4$ QCD, 
however, each quantity shows hyperscaling except for $M_\pi$, 
where $M_\pi$ obviously is outside a scaling region 
either in the conformal phase or in the chirally broken phase.
We also found that a finite-size hyperscaling fit 
with universal $\gamma$ can work by adding a mass correction term.
The fit result gives $\gamma \sim 1$ and a reasonable $\chi^2/$dof.
Further detailed analysis has been performed 
by studying an effective $\gamma$ for each hadron spectrum, 
which turns out to be useful to see a tendency towards the chiral limit. 
We found a qualitative difference between $N_f=8$ and $12$;  
in $N_f=8$ $\gamma_{\rm eff}$ for $M_\pi$ is increasing 
and seems to be approaching $\sim 1$ towards the chiral limit,  
while in $N_f=12$ a universal value of $\gamma_{\rm eff} \sim 0.4$ 
can be obtained in a smaller fermion mass region.
This result might be indicative of $N_f=8$ QCD being in the chirally broken phase.
As a result, $N_f=8$ QCD still possesses two possibilities of the strongly coupled 
conformal theory ($\gamma \sim 1$) and chirally broken theory with walking behavior.
Thus $N_f=8$ QCD is a good candidate for the walking 
technicolor model.
We have also provided a calculation of the flavor-singlet pseudoscalar mass 
in $N_f=8$ QCD. Using a gluonic operator and the gradient flow, 
we have obtained a good signal of the $\eta'$ mass for the first time in 
the many flavor QCD.
Our result suggests a heavy $\eta'$ compared to real-life QCD.

{\it Acknowledgments}
-- 
Numerical computations have been carried out on 
$\varphi$ at KMI, CX400 at the Information Technology Center in Nagoya University, 
and CX400 and HA8000 at the Research Institute for Information Technology in Kyushu University. 
This work is supported by the JSPS Grant-in-Aid for Scientific Research (S) No.22224003, (C) No.23540300 (K.Y.), 
for Young Scientists (B) No.25800139 (H.O.) and No.25800138 (T.Y.), 
and also by the MEXT Grants-in-Aid for Scientific Research on Innovative Areas No.23105708 (T.Y.)
and No.25105011 (M.K.).
This work is supported by the JLDG constructed over the SINET of NII.
The work of H.O. is supported by the RIKEN Special Postdoctoral Researcher program.
E.R. acknowledges the support of the U.S. Department of Energy 
under Contract DE-AC52- 07NA27344 (LLNL).

\end{document}